\documentclass{mem}
\usepackage{natbib}\usepackage{txfonts}\usepackage{balance}
\usepackage{graphicx}
\usepackage[a4paper,breaklinks,dvipdfm]{hyperref}
\idline{75}{282}
\begin{document}
\def\teff{$T\rm_{eff }$}
\def\kms{$\mathrm {km s}^{-1}$}
\def\zem{$z_{\rm em}$}
\def\zabs{$z_{\rm abs}$}
\def\zphot{$z_{\rm phot}$}
\def\kms{km~s$^{-1}$}
\def\l{$\lambda$}

\def\e{et~al.}
\def\ha{H-$\alpha$}
\def\hb{H-$\beta$}
\def\hg{H-$\gamma$}
\def\hd{H-$\delta$}
\def\lya{Ly-$\alpha$}
\def\lyb{Ly-$\beta$}
\def\oii{[O~{\sc ii}]}
\def\oiii{[O~{\sc iii}]}
\def\nii{[N~{\sc ii}]}
\def\sii{S~{\sc ii}}
\def\neiii{[Ne~{\sc iii}]}

\def\hi{H~{\sc i}}
\def\hii{H~{\sc ii}}

\def\nhi{\mbox{$\sc N(\sc H~{\sc I})$}}
\def\lognhi{\mbox{$\log \sc N(\sc H~{\sc I})$}}

\def\caii{Ca~{\sc ii}}
\def\caiii{Ca~{\sc iii}}
\def\ci{C~{\sc i}}
\def\cii{C~{\sc ii}}
\def\civ{C~{\sc iv}}
\def\feii{Fe~{\sc ii}}
\def\feiii{Fe~{\sc iii}}
\def\mgi{Mg~{\sc i}}
\def\mgii{Mg~{\sc ii}}
\def\mni{Mn~{\sc i}}
\def\mnii{Mn~{\sc ii}}
\def\naid{Na~{\sc ID}}
\def\siii{Si~{\sc ii}}
\def\siiii{Si~{\sc iii}}
\def\siiv{Si~{\sc iv}}
\def\tiii{Ti~{\sc ii}}
\def\niii{Ni~{\sc ii}}
\def\alii{Al~{\sc ii}}
\def\aliii{Al~{\sc iii}}
\def\znii{Zn~{\sc ii}}
\def\crii{Cr~{\sc ii}}
\def\coii{Co~{\sc ii}}
\def\oi{O~{\sc i}}
\def\ovi{O~{\sc vi}}
\def\ni{N~{\sc i}}
\def\nai{Na~{\sc i}}

\title{Nitrogen Abundances in Damped Ly$\alpha$ Absorbers}

\author{T. Zafar\inst{1}, M. Centuri\'on\inst{2}, P. Molaro\inst{2}, C. P\'eroux\inst{3}, V. D'Odorico\inst{2}, G. Vladilo\inst{2} }
 \offprints{T.\,Zafar} 

\institute{European Southern Observatory, Karl-Schwarzschildstrasse 2, 85748 Garching bei M$\ddot{u}$nchen, Germany. \email{tzafar@eso.org}
\and Osservatorio Astronomico di Trieste, Istituto Nazionale di Astrofisica, via G.B. Tiepolo 11, Trieste, Italy.
\and Aix Marseille Universit\'e, CNRS, LAM (Laboratoire d'Astrophysique de Marseille) UMR 7326, 13388, Marseille, France. }

\authorrunning{T.\,Zafar et al.}

\titlerunning{Nitrogen Abundances in DLAs}

\abstract{
Nitrogen is thought to have both primary and secondary origins depending on whether the seed carbon and oxygen are produced by the star itself (primary) or already present in the interstellar medium (secondary) from which star forms. Damped Ly$\alpha$ (DLA) and sub-DLA systems with typical metallicities of $-3.0\lesssim {\rm Z/Z}_{\odot} \lesssim-0.5$ are excellent tools to study nitrogen production. We made a search for nitrogen in the European Southern Observatory (ESO) Ultraviolet Visual Echelle Spectrograph (UVES) advanced data products (EUADP) database. In the EUADP database, we find 10 new measurements and 9 upper limits of nitrogen. We further compiled DLA/sub-DLA data from the literature with estimates available of nitrogen and $\alpha$-elements. This yields a total of 98 systems, i.e. the largest nitrogen abundance sample investigated so far. In agreement with previous studies, we indeed find a bimodal [N/$\alpha$] behaviour: three-quarter systems show a mean value of [N/$\alpha$] $=-0.87$ with a scatter of 0.21 dex and one-quarter shows ratios clustered at [N/$\alpha$] $= -1.43$ with a lower dispersion of 0.13 dex. The high [N/$\alpha$] group is consistent with the blue compact dwarves and dwarf irregular galaxies, suggesting primary nitrogen production. The low [N/$\alpha$] group is the lowest ever observed in any astrophysical site and probably provides an evidence of the primary production by fast rotating massive stars in young sites. Moreover, we find a transition between the two [N/$\alpha$] groups around [N/H] $\simeq-2.5$. The transition is not abrupt and there are a few systems lying in the transition region. Additional observations of DLAs/sub-DLAs below [N/H] $<-2.5$ would provide more clues.
\keywords{Galaxies: formation -- galaxies: evolution -- galaxies: abundances -- galaxies: ISM -- quasars: absorption lines -- intergalactic medium}
}
\maketitle{}

\section{Introduction}
Nitrogen is believed to has two production pathways either `primary' or `secondary'. Secondary production is the dominant process where production can happen in the H-burning layers of stars of all masses (mainly low and intermediate mass stars) as the seeds C and O are present in these layers. The secondary production takes place at high metallicities ([O/H] $>$ $-1$) and shows a correlation between (N/O) and (O/H). The primary production occurs at low metallicities where N goes in lockstep with O and (N/O) remains constant. Primary production can happen in intermediate mass stars (4 $\leq$ ${\rm M/M}_\odot$ $\leq$ 8) which undergo hot-bottom burning in the asymptotic giant branch (AGB) phase \citep{henry00,marigo01}.

Measurements of nitrogen have been performed in different astrophysical sites. The measurements of \hii\ regions of blue compact dwarf \citep[BCD;][]{izotov04} and dwarf-irregular galaxies \citep{vanzee06} show a primary behaviour. The nitrogen estimates in metal-poor emission line galaxies \citep{nava06,perez09} and spirals \citep{vanzee98} indicate a secondary behaviour. The abundance determination were also done in the low-metallicity galactic halo stars \citep{spite05}.

\begin{figure}[]
\resizebox{\hsize}{!}{\includegraphics[clip=true]{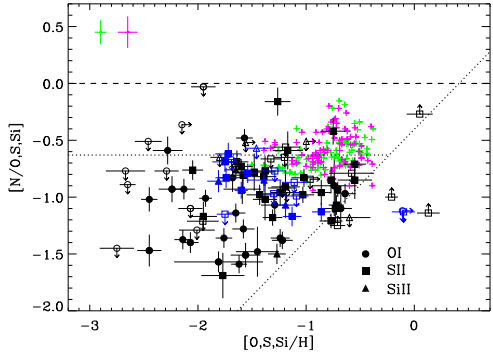}}
\caption{\footnotesize
The [N/$\alpha$] ratios against the $\alpha$-capture element metallicity. The circles, squares, and triangles indicate the O, S, and Si abundances in DLAs/sub-DLAs, respectively. The filled symbols represent measurements while open symbols show upper or lower limits. The blue symbols are the new estimates from the EUADP sample. The magenta and green pluses are measurements in the \hii\ regions of dwarf irregular \citep{vanzee06} and BCD galaxies \citep{izotov04} respectively. The magenta and green lines on the top-left corner correspond to the average errorbars for the \hii\ regions of dwarf irregular and BCD galaxies respectively. The diagonal dotted line indicates secondary production. The horizontal dotted line (primary) is plotted at the mean value of [N/O] in BCD galaxies. The dashed line indicates the solar level.}
\label{NA_OH}
\end{figure}

Chemical evolution models predict that low and intermediate mass stars can produce primary nitrogen at low metallicities. Nitrogen production by massive stars is also expected but with numerous uncertainties \citep[see][]{woosley95,meynet02,chiappini05,maeder09}. Therefore, It is important to verify the contribution of massive stars at low metallicities. In this framework, quasar Damped Ly$\alpha$ systems (DLAs with log N(\hi)$>$20.3) or sub-DLAs (log N(\hi)$>$19.0) are excellent tools to study the nitrogen production. Their typical metallicities of $-3.0$ $\leq$ ${\rm Z/Z}_\odot$ $\leq$ $-0.5$ are reaching even lower than the dwarf irregular galaxies. However, studies of nitrogen abundances require an estimate of oxygen abundance to derive (N/O) ratios. Accurate oxygen and nitrogen measurements are difficult to derive in quasar absorbers due to: $i)$ line saturation issues and $ii)$ blending of line with the Ly$\alpha$ forest. Nevertheless, a number of such measurements have been performed over the years \citep{pettini02,centurion03,petitjean08, cooke11}. Previous studies claimed that majority of the DLAs are clustered around the primary plateau. However, there exist a small population which is lying well below the primary plateau \citep{prochaska02,centurion03,petitjean08}. This dichotomy between the groups is attributed to different nitrogen enrichment histories. More details can be found in \citet{zafar14}.
 
\section{Data Sample}
The search for nitrogen is made in the European Southern Observatory (ESO) Ultra-violet Visual Echelle Spectrograph (UVES) advanced data products (EUADP) database published by \citet{zafar13,zafar13b}. The dataset has been used to search for \ni\ lines in 140 DLAs/sub-DLAs, where wavelength coverage was available for absorbers above $z_{\rm abs}>1.65$. From the non-saturated and unblended lines, we derived 9 new measurements of \ni\ together with 9 upper limits. In addition, we have analysed the DLA towards QSO\,0334$-$1612 from \citep{peroux05}. In total, the new sample is, therefore, composed of 10 measurements of \ni\ and 9 upper limits. The measurements of \ni\ are typically based on the two triplets around $\lambda1134$ and $\lambda1200$\,\AA. The fits were performed using various lines from the \ni\ triplets together with other available lines, within \texttt{MIDAS/FITLYMAN} environment \citep{fontana95}. 

For EUADP dataset, 27 measurements and 9 limits of \ni\ and $\alpha$-capture element have already been reported in the literature. Furthermore, we have made a complete reappraisal of the \ni\ measurements published in the literature to date, resulting in 43 (27 measurements, 16 limits) DLAs/sub-DLAs. This comprises a total of 64 measurements and 34 upper or lower limits of \ni\ and $\alpha$-elements.

\begin{figure}[]
\resizebox{\hsize}{!}{\includegraphics[clip=true]{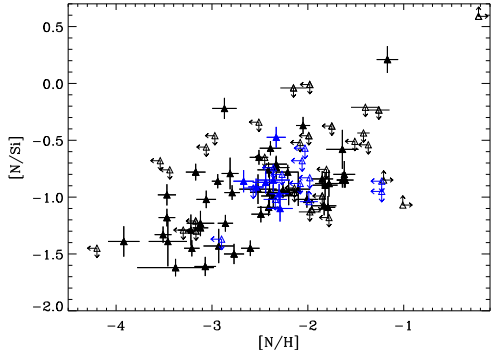}}
\caption{
\footnotesize
The [N/Si] against the nitrogen abundance in DLAs and sub-DLAs. The filled symbols indicate measurements while open ones are for upper or lower limits. The blue triangles are the new estimates from the EUADP sample.}
\label{NH_NSi}
\end{figure}

\section{The [N/$\alpha$] Bimodality}
\begin{figure}[]
\resizebox{\hsize}{!}{\includegraphics[clip=true]{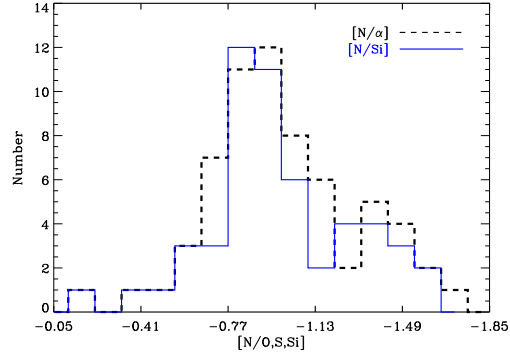}}
\caption{
\footnotesize
Histograms of [N/$\alpha$] and [N/Si] abundance ratios for DLAs and sub-DLAs in the sample. The systems where Si element is used as a proxy for [N/$\alpha$] ratios, are appearing in both histograms with the exception of 8 absorbers.}
\label{nahist}
\end{figure}

Previous studies show a bimodal distribution in the [N/$\alpha$] ratios of DLAs with a transition occurring at [N/H] $\approx$ $-3.0$ \citep{prochaska02,centurion03,petitjean08}. In Fig. \ref{NA_OH} and \ref{NH_NSi}, the bimodal distribution of [N/$\alpha$] is still apparent in the new extended sample although a transition region covering the area in between the two groups is becoming apparent. Moreover, the transition between the groups in the largest sample is found around [N/H] $\simeq-2.5$. The transition is not abrupt and there are a few systems coming in between. We find that the high [N/$\alpha$] cluster contains 76\% of DLAs/sub-DLAs (49 out of 64 measurements), showing a mean value of [N/$\alpha$] $\approx -0.87$ with a scatter of 0.21 dex. The remaining 24\% (15 out of 64) is clustered around a mean value of [N/$\alpha$] $\approx -1.43$ with a lower dispersion of 0.13 dex. The bimodality could also be seen in terms of the frequency distribution of [N/$\alpha$] ratios (see Fig.~\ref{nahist}). A Kolmogorov-Smirnov (KS) test indicates that the two populations are inconsistent at a 0.99 level of probability.

The high [N/$\alpha$] group is at the level of BCDs and dwarf irregulars \citep{izotov04,vanzee06}, suggesting that both populations experienced similar chemical evolution. In primary production N goes in lockstep with O, making [N/O] ratio constant whatever the oxygen metallicity is. Therefore, the high [N/$\alpha$] group is interpreted as due to primary N production by low and intermediate mass stars which undergo hot-bottom burning processes in the AGB phase \citep{henry00,marigo01}. Nitrogen is then ejected in the interstellar medium within a time as short as $\sim$ 50--200 Myr, since the mean-lifetimes of these stars are in that range. The large dispersion of high [N/$\alpha$] group implies that the scatter is intrinsic.

Primary nitrogen can also be produced by massive stars. Chemical evolution models predict N production either by massive stars {\it only} at high metallicities \citep{woosley95}, or by metal-poor massive stars rotating at high velocities \citep{meynet02,chiappini06}. The small dispersion in low [N/$\alpha$] group implies small or almost no intrinsic dispersion, suggesting that N production may initiated via a different process. We interpret this low [N/$\alpha$] cluster as an evidence of primary N production by fast rotating massive stars (9 $\leq$ ${\rm M/M}_\odot$ $\leq$ 20) in young sites. This could be due to the fact that intermediate mass AGB stars do not have enough time to evolve and enrich the gas with N. However, at lower timescales N production could be initiated by younger stars without invoking any change in the initial mass function. The discovery of metal-poor halo stars seems to confirm the primary N production in massive stars \citep{israelian04,spite05} with a yield that depends on stellar mass and metallicity \citep{chiappini06}.

\section{Conclusions}
Studies of nitrogen in DLAs/sub-DLAs provide important clues of the earlier stages of galactic chemical evolution. The nitrogen abundance determination in sites of low and high metallicities plays an important role in understanding the origin of nitrogen production. The largest sample used in this study, showing [N/$\alpha$] bimodal behaviour, provide clues that the enrichment mechanism needed to be different. More observations of low metallicities DLAs will help to better understand the primary N production by massive stars. 
 
\begin{acknowledgements}
TZ and CP are thankful to the BINGO! (`history of Baryons: INtergalactic medium/Galaxies cO-evolution') project by the Agence Nationale de la Recherche (ANR) under the allocation ANR-08-BLAN-0316-01. We are thankful to Ryan Cooke for providing revised abundance measurements of his data.
\end{acknowledgements}

\bibliographystyle{aa}
\bibliography{nitrogen.bib}{}

\end{document}